\documentclass[twocolumn,aps,pre,10pt,amssymb,superscriptaddress]{revtex4-1}

\usepackage{hyperref}
\usepackage{graphicx}
\usepackage{amsmath,amsfonts,amsthm,amssymb,bm}
\hypersetup{colorlinks=true, linkcolor=blue, citecolor=blue, urlcolor=blue}
\usepackage{color}
\usepackage{wrapfig}
\usepackage{enumerate}
\usepackage[version=4]{mhchem}
\usepackage[normalem]{ulem} 

\begin{document}

\title{Pair hopping in systems of strongly interacting hard-core bosons}

\author{Alvin J.R. Heng}
\affiliation{Division of Physics and Applied Physics, School of Physical and Mathematical Sciences, Nanyang Technological University, Singapore 637371, Singapore}

\author{Wenan Guo}
\affiliation{Physics Department, Beijing Normal University, Beijing 100875, China}
\affiliation{Beijing Computational Science Research Center, Beijing 100193, China}

\author{Anders W. Sandvik}
\affiliation{Department of Physics, Boston University, 590 Commonwealth Avenue, Boston, MA 02215, USA}
\affiliation{Beijing National Laboratory of Condensed Matter Physics and
Institute
of Physics, Chinese Academy of Sciences, Beijing 100190, China}

\author{Pinaki Sengupta}
\affiliation{Division of Physics and Applied Physics, School of Physical and Mathematical Sciences, Nanyang Technological University, Singapore 637371, Singapore}
\date{\today}

\begin{abstract}
We have used the Stochastic Series Expansion quantum Monte Carlo method to
study interacting hard-core
bosons on the square lattice, with pair-hopping processes supplementing
the standard single-particle hopping.
Such pair hopping arises in effective models for frustrated quantum
magnets. Our goal is to investigate the
effects of the pair hopping process on the commonly observed superfluid,
insulating (Mott), and super-solid ground-state
phases in the standard hard-core boson model with various interaction
terms. The model is specifically motivated by
the observation of finite dispersion of 2-magnon bound states in neutron
diffraction experiments \ce{SrCu_2(BO_3)_2}. Our
results show that the pair hopping has different effects on Mott phases at
different filling fractions, "melting"
them at different critical pair-hopping amplitudes. Thus, it appears that
pair hopping may have an important
role in determining which out of a potentially large number of Mott phases
(stabilized by details of the
charge-diagonal interaction terms) actually survive the totality of
quantum fluctuations present.
\end{abstract}

\maketitle

\section{Introduction}\label{sec:intro}
The interplay between competing interactions, enhanced quantum fluctuations due
to reduced dimensionality and external fields in interacting lattice bosons result 
in a rich array of novel quantum states of matter 
that have been intensely studied over the past several decades \cite{batrouni1995supersolids,batrouni2000phase,frey1997critical,hebert2001quantum,sengupta2005supersolids,jiang2012pair,schmidt2006single,freericks1994phase,kuhner1998phases,capogrosso2008monte,chen2008supersolidity}. In recent years, experimental advances have allowed the realization of these bosonic phases, such as the superfluid (SF), Bose-Einstein condensation (BEC), Mott insulator (MI) and density modulated crystalline phases with
different ordering wave vectors, in a variety of physical systems such as optical 
lattices with ultracold atoms \cite{jaksch1998cold,greiner2002quantum,morsch2006dynamics,islam2015measuring}, quantum 
magnets and excitons and polaritons \cite{kasprzak2006bose,snoke2002spontaneous} in semiconductor quantum wells. These systems are now
opening up new frontiers in the study of strongly-correlated quantum many-body systems. 

Quantum magnets, in particular, have long served as a versatile testbed for
interacting lattice bosons in a controllable manner. The low-lying magnetic
excitations, magnons, obey Bose-Einstein statistics and are an almost ideal 
realization of lattice bosons \cite{giamarchi2008bose}. The discovery of Bose-Einstein condensation in insulating magnets such as \ce{TlCuCl_3} \cite{nikuni2000bose,ruegg2003bose},
\ce{BaCuSi_2O_6} \cite{sasago1997temperature,ruegg2007multiple,jaime2004magnetic} and \ce{NiCl_2-4SC(NH_2)_2} \cite{zvyagin2007magnetic,paduan2009critical}
heralded the search for novel quantum phases of interacting bosons in quantum magnets.Often, in quantum magnets, geometrical frustration induces quantum phases and 
phenomena that are not observed in their non-frustrated counterparts, e.g., dimensional
reduction at a quantum critical point in \ce{BaCuSi_2O_6} \cite{jaime2004magnetic}
and magnetization plateaus in \ce{SrCu_2(BO_3)_2} on the geometrically frustrated 
Shastry-Sutherland lattice \cite{miyahara1999exact, kageyama1999exact, koga2000quantum}.
Understanding the nature and mechanism of formation of the
plateaus in \ce{SrCu_2(BO_3)_2} has been the subject of intense experimental 
and theoretical studies during the past 
two decades \cite{,dorier2008theory,corboz2014crystals,miyahara2000superstructures}.
The ground state of the compound is comprised of orthogonal dimer singlets
within the weakly coupled two-dimensional planes. 
In an external magnetic field, field-induced triplons constitute the lowest magnetic excitations. Theoretical
modeling and neutron scattering experiments show that strong geometric frustration significantly 
suppresses the delocalization of triplons \cite{kageyama2000direct} and
prevents the onset of field-induced
BEC of triplons that is commonly observed in other 
dimer quantum magnets
such as \ce{TlCuCl_3}. The magnetization plateaus are understood as periodic arrangements of
the triplons in regular patterns at commensurate fillings. However, the mechanism of 
triplon rearrangement into crystal orderings remain an open question. 

Several different models have been proposed in the past to 
describe the magnetization profile of \ce{SrCu_2(BO_3)_2}, treating the field-induced triplons as hard-core bosons, with varying degrees of success. These
include long range interactions and correlated nearest neighbor hopping 
of triplons, among others \cite{miyahara2003theory,momoi2000magnetization,dorier2008theory,mila2008supersolid}. 
In Ref. \cite{kageyama2000direct}, neutron scattering experiments performed by Kageyama \textit{et al.} on \ce{SrCu_2(BO3)_2} showed that while isolated triplons are localized, bound
pairs of triplons exhibit pronounced dispersion, although the cost of pair formation is high. 
This may provide a potential mechanism for the rearrangement of the triplons into 
periodic patterns observed at the magnetization plateaus and has motivated us to explore the 
role of dynamically generated triplon pairs in modifying the field driven properties
of interacting triplons. Our goal is not to derive an exact microscopic model 
of \ce{SrCu_2(BO3)_2} and provide a quantitative explanation for the magnetization
plateau formation therein. Instead, we want to isolate the effects of dispersive
bound pairs of triplons in a generic quantum magnet with multiple competing interactions through the introduction of a new effective Hamiltonian, and investigate the dynamics that arise from such a lattice model.

In this paper, we study a system of interacting hard-core bosons with single and pair hopping with nearest-neighbor (nn) and next-nearest-neighbor (nnn) repulsions on a square lattice. 
Field-induced triplons on the dimers can be faithfully mapped on to hard-core bosons through the Matsubara-Matsuda transformation \cite{matsubara1956lattice} and the dispersion of bound
pairs of triplons translate to pair-hopping processes in the bosonic model, where a pair
of hard-core bosons on nn sites hop together to the neighboring sites.
While such processes occur within the standard framework of the canonical hard-core boson
model with single-particle hopping, the amplitude of the effective process is small.
Motivated by the experimental observation in \ce{SrCu_2(BO3)_2} (suppressed single triplon
dispersion and pronounced triplon pair dispersion), the relative magnitude
of the pair hopping process is chosen to be large and considered as an independent
term in the Hamiltonian. 
Our goal here is to explore the effects of finite pair hopping on the various ground state
phases of the hard-core boson model, and to investigate if new many-body 
phases are engineered by the pair hopping process.

The paper is organized as follows. In Sec. \ref{sec:model}, we introduce our model and define the relevant parameters in the Hamiltonian. Section \ref{sec:sse} describes how the pair hopping process can be incorporated into the Stochastic Series Expansion Quantum Monte Carlo scheme, which involves a straightforward generalization from 2-site bond operators to 4-site plaquette operators. Section \ref{sec:ob} defines the observables that are measured from the Monte Carlo simulation. Section \ref{sec:results} presents the main results of our numerical simulations, where we include illustrative phase diagrams for a wide range of Hamiltonian parameters as well as more detailed observable plots and accompanying analyses. Finally, in Sec. \ref{sec:dc} we discuss the significance of our results; namely how pair hopping modifies the formation and stability of the various Mott phases and its implications to our understanding of \ce{SrCu_2(BO_3)_2}.

\section{Model}\label{sec:model}
\begin{figure}
\centering
  \includegraphics[trim=0.2cm 0.2cm 0.2cm 0.2cm, clip, width=\columnwidth]{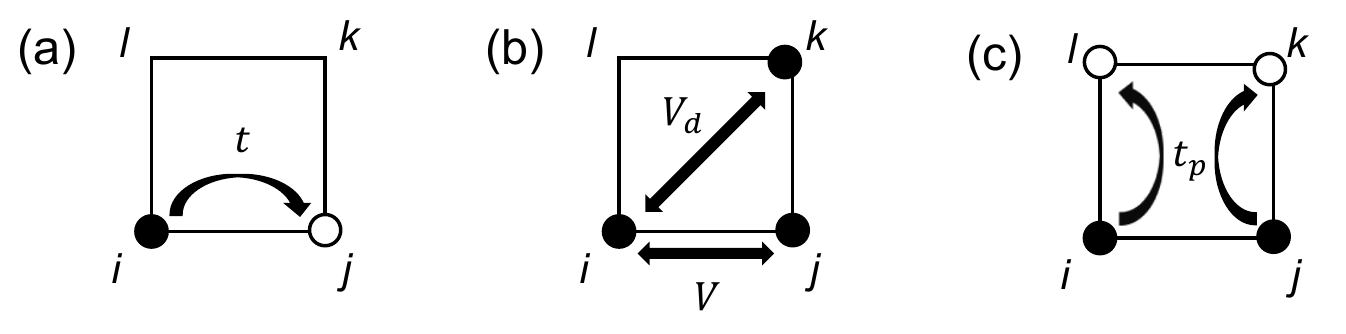}
  \caption{Illustration of the Hamiltonian parameters of Eq.~\eqref{eq:hamil} on the unit plaquette. (a) Single-particle hopping parameter $t$, (b) 
nn repulsion $V$ and nnn repulsion $V_d$, and (c) the pair-hopping parameter $t_p$. Filled circles represent the bosons while empty 
circles represent sites that a boson can hop onto.}
  \label{fig:UnpairedPaired}
\end{figure}

The Hamiltonian for the model described above is given by 
\begin{equation}\label{eq:hamil}
\begin{split}
 {\cal H} = & -t\sum_{\langle i,j\rangle}(a_i^\dagger a_j + h.c.) 
 - t_p\sum_{\square}(a_i^\dagger a_j^\dagger a_k a_l+h.c. ) \\ 
 & +V\sum_{\langle i,j\rangle}n_i n_j+V_d\sum_{\langle i,k \rangle}n_i n_k -\mu\sum_i n_i\
\end{split}
\end{equation}
where $a_i^\dagger$ and $a_j$ are the creation and annihilation operators respectively on sites $i$ and $j$. The $\square$ denotes a four-site plaquette on which our Hamiltonian parameters are defined, with sites we label $i,j,k,l$, as shown in Fig. \ref{fig:UnpairedPaired}. $n_i=a^\dagger_i a_i$ is the number operator at site $i$. $t$ and $t_p$ are the single and pair hopping amplitudes, respectively. $V$ and $V_d$ are the nn and nnn repulsion, respectively, and $\mu$ is the chemical potential. 
We work in the hard-core boson limit, i.e., the possible local occupancies are restricted to $n_i\in$ \{0,1\}. A square lattice with periodic boundary conditions of $N=L \times L$ sites is assumed. We set $t=1$ henceforth. 

The nnn repulsion term $V_d$ in Eq. \eqref{eq:hamil} is necessary for promoting pair formation. 
In Fig.~\ref{fig:UnpairedPaired}(b), bosons on diagonal sites $\langle i,k \rangle$ incur an energy cost $+V_d$, which 
increases the likelihood of nearest-neighbour pairs occuring ($\langle i,j \rangle$ in Fig.~\ref{fig:UnpairedPaired}(c)). 
The nn bosons are subsequently able to hop as pairs in proportion to the magnitude of $t_p$.

\section{Stochastic Series Expansion Method}\label{sec:sse}

We have used the Stochastic Series Expansion (SSE) Quantum Monte Carlo (QMC) \cite{sandvik1999stochastic,sandvik1991quantum} method to 
simulate the Hamiltonian (Eq. (\ref{eq:hamil})) on finite-size systems. The SSE is a finite-temperature algorithm 
based on the stochastic evaluation of the diagonal matrix elements of the density matrix, 
$\exp(-\beta {\cal H})$, in a Taylor series expansion. 

The SSE method employs the operator loop update method in sampling the configuration state space for the ground state configuration. The loop update involves the construction of a linked vertex list, where lattice sites are propagated in imaginary time, with diagonal and off-diagonal operators acting between the propagation levels according to a stored operator string. Sites connected by an operator between propagation levels are known as vertices. Configuration updates are achieved by the introduction of a `defect' - a boson occupation inversion in the hard-core limit - into a random leg of a vertex. The defect is then propagated throughout the linked list, until the defect meets its initial introduction site and the loop is closed. The propagation of the defect is stochastically sampled in a manner proportional to the weights of the resulting vertices. After closing the loop, the lattice configuration and operator string are updated to reflect the changes made.

On 2D square lattices, the SSE loop update algorithm considers operators acting on 2-site bonds, such that vertices are 4-legged: 2 sites before and 2 sites after the action of an operator. To incorporate the pair hopping procedure, one needs to consider operators beyond 2-site bond operators. In particular, we consider operators that act on the 4-site plaquette mentioned in Fig. \ref{fig:UnpairedPaired}. This means that vertices in the linked list are now 8-legged: 4 sites before and 4 sites after the action of the operator. Conventional diagonal and single particle hopping operators carry over easily to the plaquette case. Our focus of the discussion will be on the incorporation of pair hopping operators in the loop update procedure.

\begin{figure}
\centering
  \includegraphics[trim=0.0cm 0.2cm 0.0cm 0.2cm, clip, width=\columnwidth]{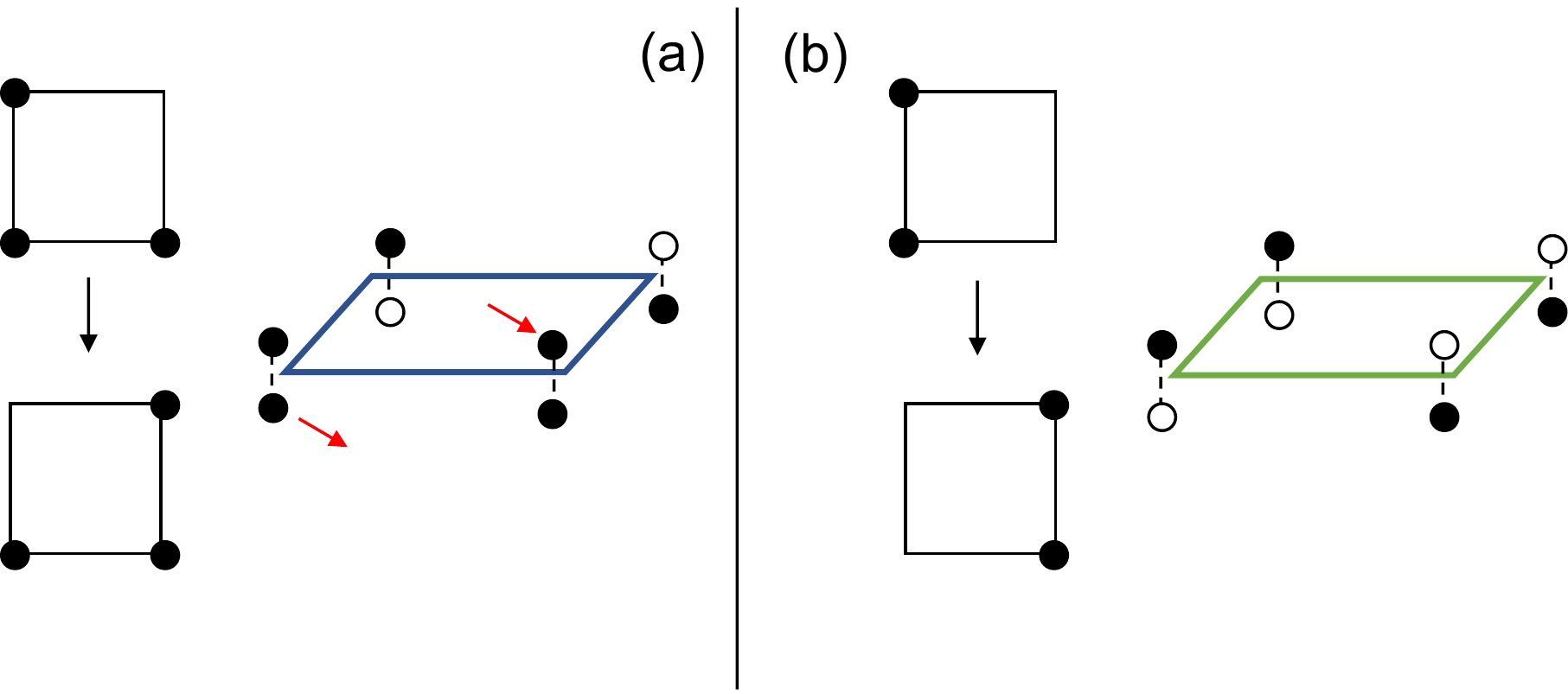}
  \caption{An illustration of the creation of a pair hopping operator through a loop update. The operator string propagation level propagates downwards in this figure. (a) The left shows a single hop operator in real space as the lattice is propagated in imaginary time. The right shows the 3D representation of a 8-legged vertex, with a single hop operator represented by a blue rectangle. Filled circles represent a boson and empty circles represent empty sites. The red arrows indicate the introduction of a defect into and out of the vertex as part of the loop update. Propagation of defects cause a boson occupancy inversion.  (b) The left shows the resulting pair hop operator acting in real space. The right shows the pair hop operator as a result of the loop update, represented by a green rectangle. Note that to obtain the pair hop operator, we simply flip the boson occupancy of the two sites indicated by red arrows in (a), which was initially a single-hop operator. This occupancy flip resulted in the conversion from a single-hop operator to a pair-hop operator.}
  \label{fig:loop_upd}
\end{figure}

We note that only slight modifications are required to achieve pair hopping in the SSE framework in the context of plaquette operators. 
Similar to the case of the hard-core boson model with `pair rotation' of two bosons occupying diagonally opposite corners of a plaquette
\cite{sandvik02jk}, the pair hop operators are introduced to the operator string with the same linked list loop update procedure that is conventional in a 2-site bond operator SSE scheme. An illustration of the procedure is shown in Fig.~\ref{fig:loop_upd}. The crucial insight comes from the fact that pair hop operators are created only from single hop operators in the loop update. Figure \ref{fig:loop_upd} shows one way in which an existing single hopping operator in the operator string can be converted to a pair hopping operator through propagation of a single defect. Consequently, this implies that in our SSE framework, a non-zero $t$ in the Hamiltonian of Eq.~(\ref{eq:hamil}) is necessary for the simulation to incorporate pair boson propagation.

\section{Observables}\label{sec:ob}

In this section, we define the observables measured in our simulations that will be the basis of our analysis in Sec. \ref{sec:results}.
The average boson density is defined as
\begin{equation}
	\langle n \rangle = \frac{1}{N}\sum_i  n_i.
\end{equation}
The total superfluid density (stiffness) is given by 
\begin{equation}
	\rho_s=\frac{\partial^2 f(\phi)}{\partial \phi^2},
    \label{eq:stiffderiv}
\end{equation}
where $f(\phi)$ is the free-energy (or ground-state energy at $T=0$) density in the presence of a phase twist $\phi$. It is
evaluated in SSE simulations as
\begin{equation}
	\rho_s=(\omega_x^2+\omega_y^2)/\beta,
    \label{eq:totstiffness}
\end{equation}
where $\omega$ is the winding number in the $x$ or $y$ directions, defined as
\begin{equation}
    \omega_\alpha = (N^+_\alpha - N^-_\alpha)/L, \ \ \ (\alpha=x,y).
\end{equation}
$N_\alpha^+$ is the total number of particle hops in the arbitrarily chosen positive direction of the lattice. This implies a pair particle hop in the positive direction increments $N_\alpha^+$ by 2. On a conventional Bose-Hubbard model without pair hopping, it is identical to the total number of operators $a_i a^\dagger_j$ in the QMC operator string, if site $j$ is in the positive direction of site $i$.

To quantify the magnitudes of single and pair particle hopping separately, we define the single and pair particle stiffness, $\rho_t$ and $\rho_{tp}$, respectively, as
\begin{equation}
	\rho_{t} = (\omega_{t,x}^2+\omega_{t,y}^2)/\beta,
\end{equation}
and
\begin{equation}
	\rho_{tp} = (\omega_{tp,x}^2+\omega_{tp,y}^2)/\beta.
\end{equation}
$\omega_{t,\alpha}$ and $\omega_{tp,\alpha}$ are the net sum of the single and pair particle hops, respectively, for $\alpha=x,y$. Concretely, we define them as
\begin{equation}\label{eq:ttpwinding}
\begin{split}
	\omega_{t,\alpha} &= (N^+_{t,\alpha} - N^-_{t,\alpha})/L, \\
   	\omega_{tp,\alpha} &= (N^+_{tp,\alpha} - N^-_{tp,\alpha})/L,
\end{split}
\end{equation}
where $N^+_{t,\alpha}$ is the total number of single particle hops in the positive direction, and $N^+_{tp,\alpha}$ is the total number of pair hops in the positive direction. A pair  hop in the positive direction increments $N^+_{tp,\alpha}$ by 2, and vice-versa. From our definitions in Eq. \eqref{eq:ttpwinding}, 
\begin{equation}
\omega_{\alpha}=\omega_{t,\alpha }+\omega_{
    tp,\alpha}, \ \ \ (\alpha=x,y)
    \label{eq:totwinding}
\end{equation} 
i.e., the total winding number is the sum of the single and pair winding numbers. Note that due to the way the various stiffness are defined,
\begin{equation}
\rho_s \neq \rho_{t} + \rho_{tp}.
\end{equation}
It should be noted that $\rho_t$ and $\rho_{tp}$ serve as useful quantities in measuring the relative contributions of single and pair currents in the system, but do not constitute experimentally measurable observables such as the total stiffness $\rho_s$ defined in Eq.~(\ref{eq:stiffderiv}).

In order to identify the presence of density modulation, or equivalently crystal ordering, we compute the static structure factor 
\begin{equation}
    S(\vec{k})=\frac{1}{N^2}\sum_{\vec{r}} e^{i\vec{k} \cdot \vec{r}}C(i,j),
\end{equation}
where $\vec{r}$ is the vector representing the separation of sites $\langle i,j \rangle$, and $\vec{k}=(k_1,k_2)$ is the wave vector, where $k_1,k_2 \in [0,2\pi]$. $C(i,j)$ is the density-density correlation function \cite{sandvik1997finite}, defined as 
\begin{equation}
	C(i,j) = \langle n_i n_j \rangle.
\end{equation}
Simulations in this study are done at $\beta=L$ to extract ground-state properties, with simulated annealing \cite{kirkpatrick1983optimization} carried out at the equilibration step of the operator string to ensure convergence of the QMC simulation.

\section{Results}\label{sec:results}

\begin{figure}
\centering
  \includegraphics[trim=0.0cm 0.5cm 0.0cm 0.0cm, clip,  width=\columnwidth]{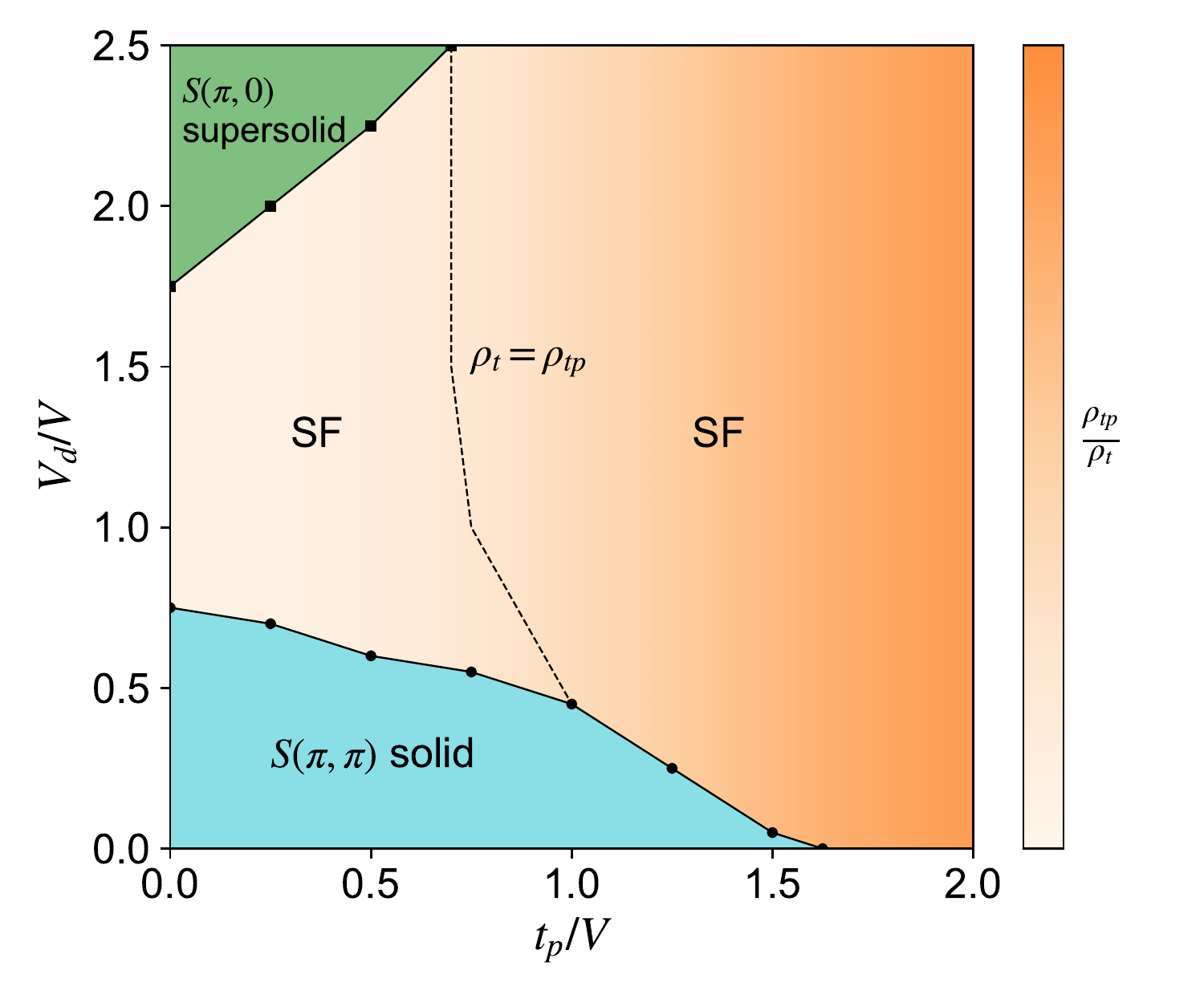}
  \caption{Ground-state phase diagram of Eq. \eqref{eq:hamil} at $t/V=1/4, \mu/V=5$. Lines are guide to the eyes. The dotted line indicates the boundary where $\rho_t = \rho_{tp}$. The orange intensity represents the magnitude of the ratio $\rho_{tp}/\rho_t$, which increases with $t_p$. A $S(\pi,0)$ solid is stabilized at larger $\mu/V$ and is not shown in this phase diagram.}
  \label{fig:phasediagram2}
\end{figure}

A representative ground-state phase diagram of the model, Eq.~\eqref{eq:hamil}, in 
the parameter space of the pair-hopping amplitude, $t_p$, and the strength of 
next-nearest neighbor interaction, $V_d$, at fixed $t$ (single-particle hopping 
amplitude) and $\mu$ (the chemical potential) is shown in Fig. \ref{fig:phasediagram2}. 
The nearest neighbor interaction strength, $V$, is chosen as the unit of energy 
and the Hamiltonian parameters are expressed in units of $V$. In the limit of
$t_p=0=V_d$, Eq. \eqref{eq:hamil} reduces to the 
canonical Bose Hubbard model where the ground state for the chosen values of $t$ 
and $\mu$ is a checkerboard solid, with an ordering
wave vector $\bf{k} = (\pi,\pi)$. The density of particles is constant at 
$\langle n \rangle = \frac{1}{2}$ and there is a gap to adding or removing a boson.
As the strength of the next-nearest neighbor interaction is increased (at $t_p = 0$), 
there is a transition to the superfluid phase at an intermediate value of $V_d/V$,
where competing nn and nnn interactions suppress any crystallization
of the bosons into a density wave. 
Eventually, for sufficiently strong nnn neighbor repulsion, the ground
state enters a supersolid (SS) phase. The wave vector  of the
underlying solid order (density modulation of the bosons) changes to $(\pi,0)$,
reflecting a striped solid. The density of particles deviates from 
$\langle n \rangle = \frac{1}{2}$ and the additional particles form a superfluid
that co-exists with the solid ordering, resulting in a SS ground state. The 
pair-hopping process enhances the extent of superfluid phase at the cost of 
the solid orders, suppressing both the checkerboard solid and SS phases completely
for sufficiently strong $t_p$. The SF phase has contributions from both single
particle and pair currents -- this is confirmed by finite values of the stiffness
for both currents, viz., $\rho_t$ and $\rho_{tp}$. The pair current contribution
is finite for {\it any} non-zero $t_p$, with the relative contribution increasing 
monotonically with $t_p$ (as shown by the color density profile in the phase 
diagram) \footnote{It is important to note that although the Hamiltonian (Eq. (\ref{eq:hamil}))
conserves both the particle number and the number of pairs, in the simulations, only the 
total current is conserved in lieu of the individual $t$ and $t_p$ currents.
Consequently, one gets $\rho_t,\rho_{tp} > 0$, even in
the insulating phases, where the currents should be zero. This is due to Gaussian 
fluctuations of the individual currents regardless of winding. Hence we observe small, 
non-zero $\rho_{tp}$ even in the Mott phases. The insulating character of the phases
is confirmed by the vanishing of the total stiffness. It should be pointed
out that the individual stiffness values become much larger in the non-Mott phases,
along with finite value of the total stiffness.}.
\begin{figure}
\centering
  \includegraphics[trim=0.0cm 0.5cm 0.0cm 0.0cm, clip,  width=\columnwidth]{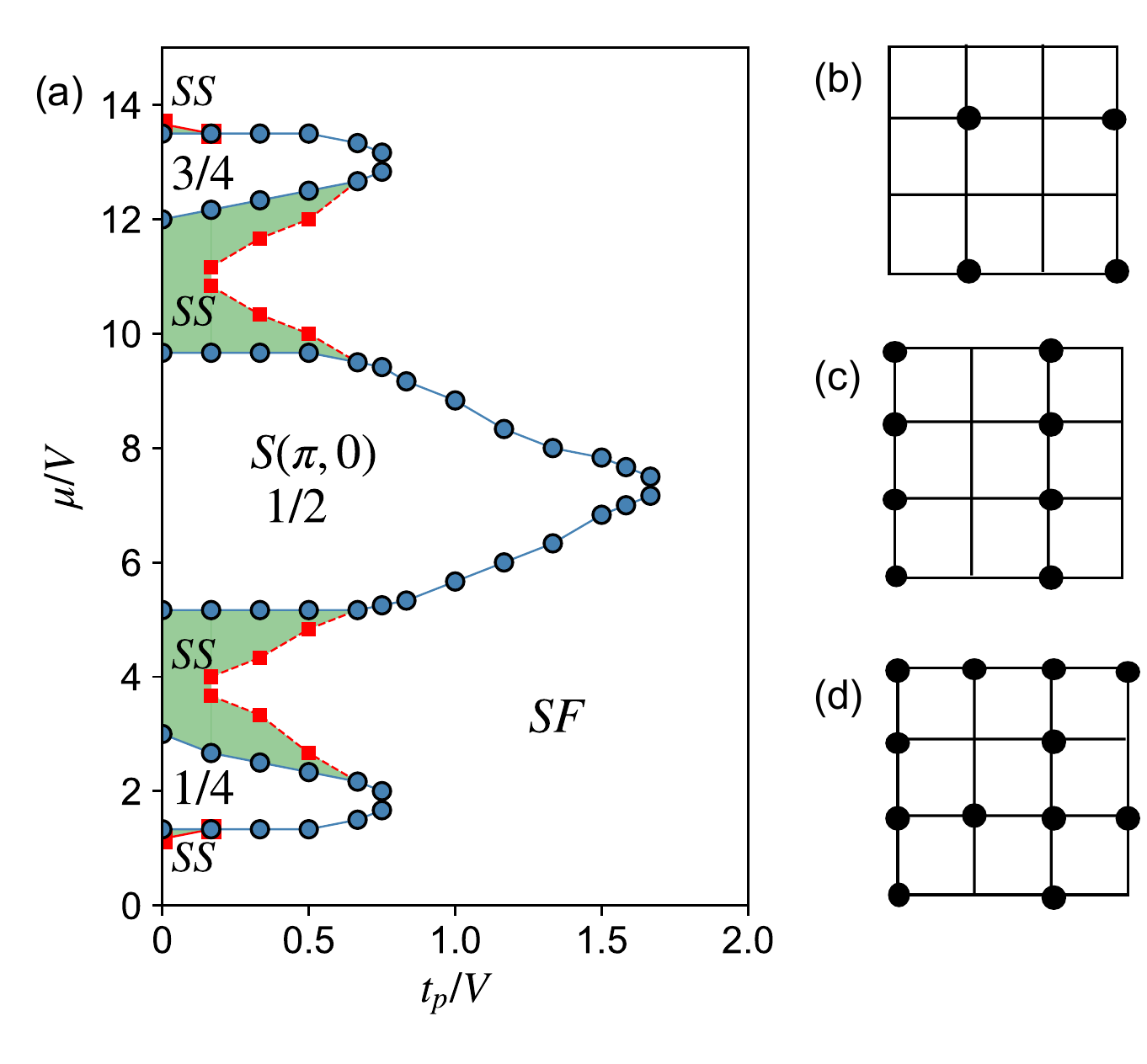}
  \caption{(a) Ground-state phase diagram of Eq. \eqref{eq:hamil} with $t/V=1/6$ and $V_d/V = 5/3$. Three distinct Mott insulating lobes are present with their densities indicated, as well as supersolid and superfluid phases. Illustrations of the ground state configuration of the (b) $\frac{1}{4}$, (c) $\frac{1}{2}$ and (d) $\frac{3}{4}$ solids on a $L=4$ lattice.}
  \label{fig:phasediagram1}
\end{figure}

In our model,
the transitions between various density wave phases are modified by the
appearance of intervening supersolid phases. This is aptly 
demonstrated in the ground-state phase diagram in the parameter space of the 
pair-hopping amplitude, $t_p$, and the chemical potential, $\mu$, at fixed $t$ , 
$V=6t$ (the nn repulsion) and $V_d=5/3V$ (the nnn repulsion), shown in Fig. \ref{fig:phasediagram1}(a). 
The pair hopping and chemical potential are expressed in units of $V$. Three distinct Mott insulating lobes are present, corresponding to different filling factors, as the chemical potential $\mu$ is varied. The  solid phases are destabilized with an increasing $t_p$, as the large pair-hopping amplitude suppresses any crystallization of the bosons into a density wave. This is manifested by the predominantly SF character of the ground state at large $t_p$. At sufficiently low $t_p$ and $\mu$, the system is in a gapless SF phase, with zero energy cost to the addition of a boson. With increasing $\mu$, there is a transition into a $\frac{1}{4}$ solid phase, which is characterized by a vanishing stiffness and finite gap. The bosons form a density wave with a pattern schematically shown in Fig. \ref{fig:phasediagram1}(b).  Increasing $\mu$, the system undergoes a transition to a SS phase, which is characterized by a finite solid order and co-existing superfluid density. Due to $V_d>V$, the wave vector of the underlying solid order is $(\pi,0)$, reflecting a striped solid. With further increase in $\mu$, a discontinuous transition drives the ground state to a $\frac{1}{2}$ solid, where the nnn repulsion crystallizes the bosons into alternating stripes, as shown in Fig. \ref{fig:phasediagram1}(c). Finally, another SS phase with $(\pi,0)$ ordering separates the $\frac{1}{2}$ solid and the $\frac{3}{4}$ solid at large values of $\mu$. The boson ordering of the $\frac{3}{4}$ solid is shown in Fig. \ref{fig:phasediagram1}(d).

Significantly, no new phases -- such as additional density wave
phases -- are stabilized by the introduction of the pair-hopping process. The boson ordering of the 
solid phases remain unchanged by pair-hopping as well. This is in 
contrast to the case of the hard-core boson model with 'pair rotation', which flips two bosons residing on 
opposite diagonal corners of a plaquette to the other diagonal on the same plaquette. In the mentioned model, even without any
diagonal interaction terms, the two-body kinetic term can induce new solid phases 
\cite{sandvik02jk,sandvik06annals}.

A key observation from Fig. \ref{fig:phasediagram1} is that the
different Mott lobes are modified differently by the pair hopping process from their 
counterparts when only $t$ is present. This
is nicely illustrated by the observation that the $\langle n \rangle = \frac{1}{2}$ 
lobe is significantly larger than the $\frac{1}{4}$ and $\frac{3}{4}$ lobes, and 
persists in a larger range of $t_p$ and $\mu$. This has important implications in 
realistic models with long range interactions. While the $t$-only model may exhibit 
several plateaus, their extent will be heavily modified by any pair-hopping process, including the possible suppression of some of them. Another interesting feature is 
that {\it all} the transitions into and out of the Mott phases are discontinuous in 
nature. This is analogous to meta-magnetism in spin models \cite{iaizzi2017field,iaizzi2018metamagnetism} and we plan to investigate this further in future studies.

\begin{figure}
\centering
  \includegraphics[trim=0.4cm 0.4cm 0.4cm 0.4cm, clip, width=\columnwidth]{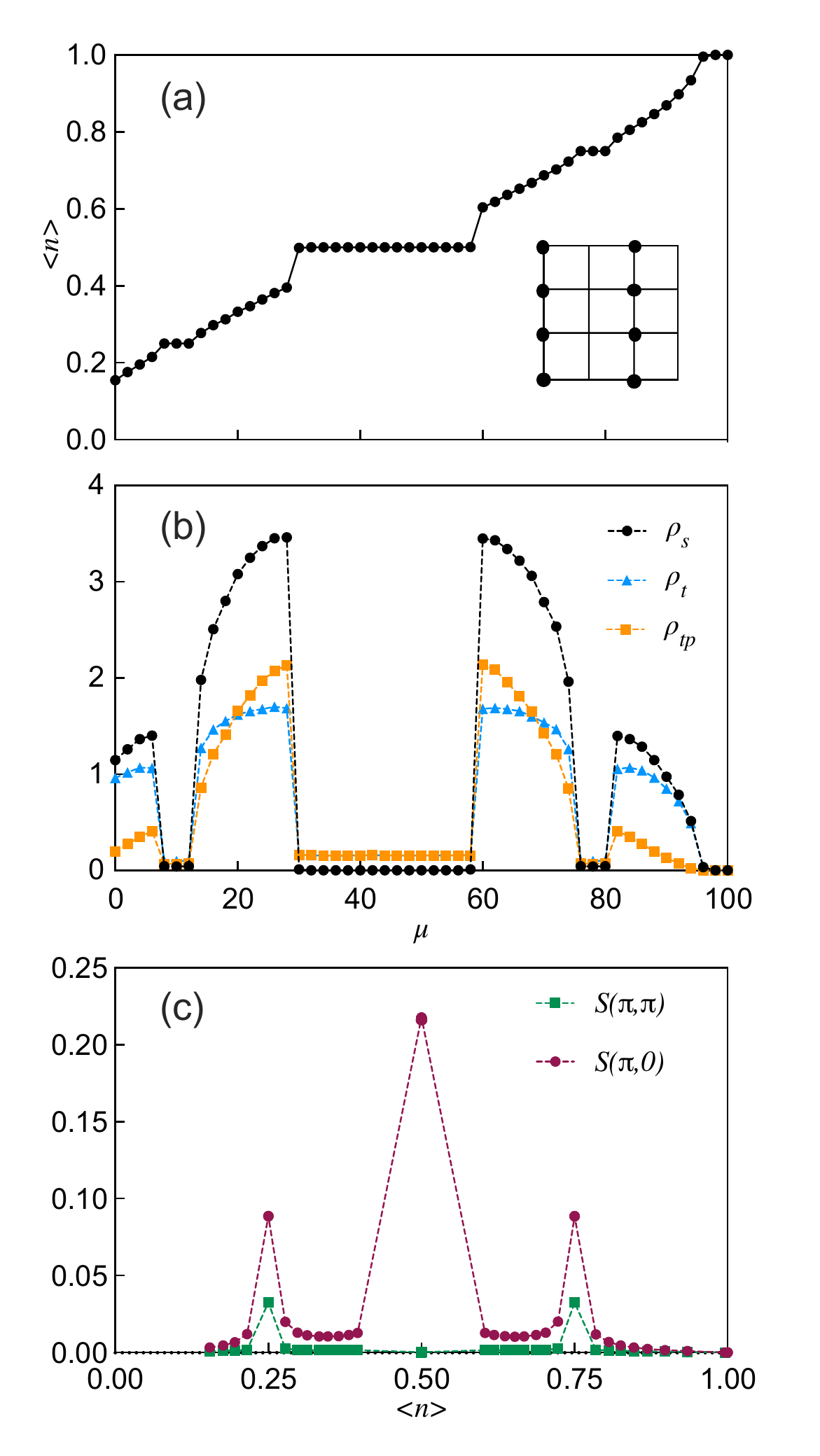}
  \caption{(a) Plot of boson density $\langle n \rangle$, (b) plot of total stiffness $\rho_s$, single stiffness $\rho_t$ and pair stiffness $\rho_{tp}$, (c) plot of $S(\pi,\pi)$ and $S(\pi,0)$. Parameters for this simulation are $t_p=4, V=6, V_d=10$ for a $L=12$ system.}
  \label{fig:1}
\end{figure}

Magnetization plateaus in spin models manifest as boson density plateaus in the boson model. To demonstrate the dynamics of pair hopping on the density plateaus, we plot the full range of observables in Fig. \ref{fig:1}, with parameters equivalent to taking a slice of constant $t_p=4$ in the phase diagram of Fig. \ref{fig:phasediagram1}(a). For the parameters chosen, we observe the existence of the aforementioned $\langle n \rangle =\frac{1}{4}$, $\frac{1}{2}$ and $\frac{3}{4}$ density plateaus in Fig. \ref{fig:1}(a). We note that the $\langle n \rangle =\frac{1}{4}$, $\frac{1}{2}$ plateaus correspond to the $m/m_s=\frac{1}{4}$ and $\frac{1}{2}$ plateaus proposed by other studies \cite{miyahara2000superstructures}. Discontinuities in the first derivative of the density and total stiffness, $\rho_s$, indicate discontinuous phase transitions into and out of the three solid phases. 

To study the solid ordering in the various plateaus, we plot the structure factor $S(\pi,\pi)$ and $S(\pi,0)$ as a function of $\langle n \rangle$ in Fig. \ref{fig:1}(c). A finite $S(\pi,\pi)$ corresponds to a checkerboard boson ordering, while a finite $S(\pi,0)$ corresponds to striped boson ordering. As we have set $V_d>V$ in this simulation, the striped ordering out competes the checkerboard ordering and we observe a striped solid at $\frac{1}{2}$-filling, characterized by a peaked $S(\pi,0)$. The total stiffness vanishes in this phase, demonstrating the gapped nature of the striped solid, where it is energetically prohibitive to add another boson. 

Compared to the $ \langle n \rangle=\frac{1}{2}$ solid, the situation is markedly different for $\langle n \rangle = \frac{1}{4}$ and $\frac{3}{4}$. The $\frac{1}{4}$ solid is stabilized by bosons avoiding both nn ($V$) and  nnn ($V_d$) repulsive interactions \cite{dang2008vacancy}, which is obvious in Fig. \ref{fig:phasediagram1}(b). The $\frac{1}{4}$ solid is then gapped as the addition of one boson incurs energy costs of either $2V-\mu$ or $4V_d-\mu$, depending on the neighborhood configuration of the site chosen. On the other hand, the $\frac{3}{4}$ solid manifests as a sequence of alternating fully-filled and half-filled stripes, as shown in Fig. \ref{fig:phasediagram1}(d). It is clear from the figures that the two phases are related by a particle-hole symmetry. Again, the gapped nature of the $\frac{3}{4}$ solid is obvious, as the addition of a boson incurs an energy cost of $4V+4V_d-\mu$. The gapped nature of both phases is also evident by the vanishing stiffness shown in Fig. \ref{fig:1}.

\begin{figure}
\centering
  \includegraphics[trim=0.2cm 0.2cm 0.2cm 0.2cm, clip, width=\columnwidth]{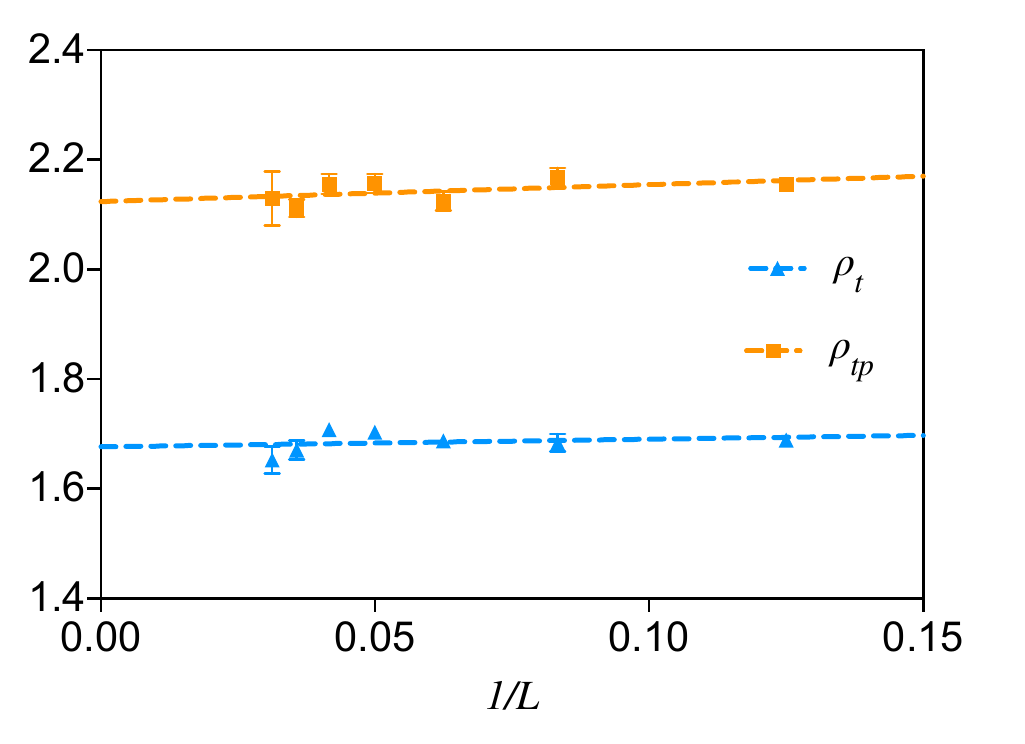}
  \caption{Finite-size scaling of $\rho_{t}$ and $\rho_{tp}$ as a function of the inverse system size $1/L$. The chemical potential is fixed as $\mu=28$ for all data points such that the system is in the SF phase. Other Hamiltonian parameters are identical to Fig. \ref{fig:1}.}
  \label{fig:fss}
\end{figure}

To characterize the relative magnitudes of single and pair particle flow, we plot 
$\rho_t$ and $\rho_{tp}$ separately in Fig. \ref{fig:1}(b). We note that single and pair particle flow 
co-exist at all values of $\mu$ for non-zero $t$ and $t_p$. The two currents reinforce 
each other in the SF phase, resulting in
a total stiffness which is greater than the individual contributions from the
single particle and pair currents. One could also have counter-propagation that would
cause partial cancellation of the currents and a smaller $\rho_s$
than $\rho_t$ and $\rho_{tp}$. Here, we observe this effect in the $\frac{1}{2}$ striped solid, where $\rho_t=\rho_{tp}$ are non-zero, yet $\rho_s$ vanishes. The origin of this counter-propagation is due to trivial local fluctuations in the single and pair currents, such that in the solid phase, two single boson hop fluctuations that break the staggered density pattern is exactly cancelled by a pair boson hop in the opposite direction, and vice-versa. This effectively conserves a vanishing $\rho_s$ in the solid, even while $\rho_t$ and $\rho_{tp}$ are non-zero.

The stiffness plots exhibit a reflection symmetry about the $\frac{1}{2}$ solid. At small and large filling factors, $\rho_t$ is larger than $\rho_{tp}$, despite the fact that $t_p = 4t$. This is explained as at low fillings, boson occupancy is sparse, making it unlikely for bosons to meet as nn pairs. At large filling factors, the lattice becomes crowded and the presence of pairs of holes, such that boson pairs can hop to fill the holes, become unlikely. This results in a larger $\rho_t$ than $\rho_{tp}$, despite the significantly larger pair hopping amplitude $t_p$. It is at intermediate filling factors where $\rho_{tp}>\rho_t$, in a phase we call 'pair superfluidity'. Intermediate filling factors satisfy the conditions that the lattice is neither too sparse or too crowded, thus being conducive for pair boson hopping. A finite-size scaling analysis of $\rho_{tp}$ in Fig. \ref{fig:fss} shows that pair superfluidity is not merely a finite-size effect, and the phase extends to the thermodynamic limit. Additionally, we find that by varying the Hamiltonian parameters for a large range of values (not shown), pair superfluidity is achieved only when we tune $t_p\gtrapprox 4t$.

\begin{figure}
\centering
  \includegraphics[trim=0.2cm 0.4cm 0.2cm 0.2cm, clip, width=\columnwidth]{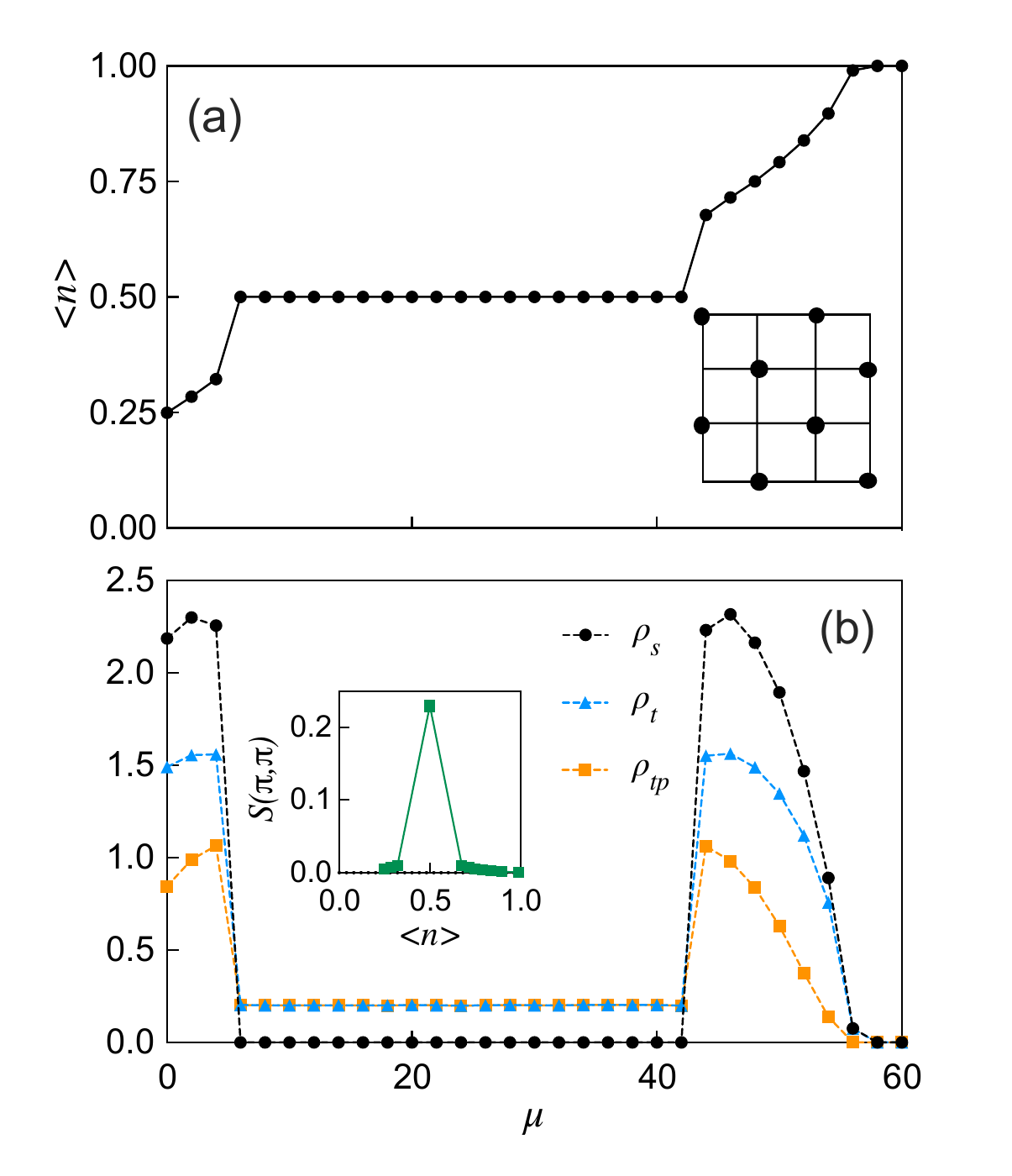}
  \caption{The boson density (a) and the various stiffness constants (b) as functions of $\mu$. The inset of (a) illustrates the geometry of bosons at the $\langle n \rangle = \frac{1}{2}$ checkerboard plateau. The inset of (b) plots $S(\pi,\pi)$ as a function of boson density. Parameters for this simulation are $t_p=4, V=6, V_d=0$ for an $L=8$ system.}
  \label{fig:2}
\end{figure}

A parameter set that stabilizes a checkerboard solid at $\frac{1}{2}$ filling is shown in Fig. \ref{fig:2}. The absence of the $\frac{1}{4}$ and $\frac{3}{4}$ plateaus in the density plot are a result of the lack of simultaneous nn and nnn repulsion, which as mentioned are necessary in the formation of these solid phases. However, a checkerboard solid at $\frac{1}{2}$ filling is still stabilized, characterized by a strongly peaked $S(\pi,\pi)$. 

In this simulation, we notice that despite having $t_p/t=4$, as with the simulation in Fig. \ref{fig:1}, $\rho_{tp}$ is significantly smaller than $\rho_t$ in the SF phases. Importantly, in the SF phase, $\rho_{tp}<\rho_t$ for \textit{all} $\mu$ points, even at intermediate SF filling factors where it was mentioned to be the most favorable for boson pair hopping. This observation is due to the difference in boson ordering for parameters that stabilize a checkerboard and striped solid at half filling. In the SF phase of the former case, bosons will still satisfy a checkerboard ordering as far as possible to minimize nn repulsions. In a checkerboard-like configuration, bosons are largely not occupying nn sites and pair-hopping of bosons then becomes impossible, despite a large $t_p$. This results in $\rho_{tp}$ being significantly suppressed. In the latter case, a striped-like ordering in the SF phase implies that bosons are largely paired up, allowing pair hopping of bosons to occur more frequently. This results in more dispersive pair hopping of bosons, and subsequently a larger $\rho_{tp}$. Therefore, a dispersive pair hopping of bosons is stabilized by large $t_p$ \textit{and} $V_d$ in the Hamiltonian of Eq. \ref{eq:hamil}. Incidentally, we observe the same effects of currents counter-propagation in the $\frac{1}{2}$ checkerboard solid as we did in the striped solid in Fig. \ref{fig:1}, as evident by $\rho_t=\rho_{tp}$. The mechanism by which counter-propagation manifests in the checkerboard solid is identical to that of the striped solid.

\section{Discussion and Conclusion}\label{sec:dc}
Our results provide useful insight into the role of pair hopping process on
the ground state phases of a system of interacting hard-core bosons. 
As discussed earlier, the microscopic origin of 
magnetization plateaus in the frustrated quantum magnet \ce{SrCu_2(BO3)_2} 
remains incompletely understood.
Neutron scattering experiments show that single magnon excitations are almost completely
dispersionless. Naively, one might expect this to result in a glassy dynamics in the
presence of a magnetic field. Interestingly, the same neutron scattering experiments 
reveal that bound states of two magnons have pronounced dispersion, although the
cost of formation of such pairs is high. This provides a potential mechanism for
the delocalization of field-induced triplons necessary for the long ranged
ordering of the triplons at the magnetization plateaus. However, important questions 
remain: does the dispersion of bound pairs retain the stability of the plateaus? Do
they result in new processes that are inconsistent with experimental observations in
\ce{SrCu_2(BO3)_2}? While this is observed experimentally, such a pair hopping term has
never been incorporated into any model hamiltonian describing \ce{SrCu_2(BO3)_2}. As 
such, a rigorous microscopic simulation such as ours studying the effects
of such a process is valuable. Our Hamiltonian, Eq.~\eqref{eq:hamil}, mimics the dispersion
of bound pairs as a pair-hopping process. In keeping with the experimental observations, the
amplitude of the pair hopping process is chosen to be much greater than the single
particle hopping process. The high energy of formation is reflected in the finite near-neighbor
repulsion, $V$. Our results demonstrate conclusively that highly dispersive
magnon bound pairs are compatible with the formation of magnetization plateaus.
However, the exact sequence of plateaus observed in \ce{SrCu_2(BO3)_2} is different
from the results obtained here. It is highly plausible that one needs to introduce  longer-range interactions, beyond what a 4-site plaquette QMC scheme can accommodate, to fully obtain the plateaus observed in \ce{SrCu_2(BO3)_2}.
Hence, while our results do not provide a comprehensive understanding of
{\it all} plateaus in the experimental system, it provides a plausible explanation
for their formation mechanism in the absence of any significant single magnon 
dispersion. We have also demonstrated that $t_p$
is important in governing which plateaus actually survive. In
principle one might have a huge number of plateaus for realistic
interactions with only $t$, but $t_p$ has different effects on different
Mott phases and some of them will be destroyed by $t_p$ even though
they survive in the presence of $t$ only. 

In summary, we have investigated the
role of a pair hopping process on the ground state phases of interacting hard
core-bosons on a 2D square lattice. Our results may provide useful insight into the mechanism of delocalization 
of field-induced triplons in the frustrated magnet, \ce{SrCu_2(BO3)_2}, necessary for
the formation of periodic patterns observed in the magnetization plateaus.

\begin{acknowledgments}
A.J.R.H. thanks the CN Yang Scholars Programme for financial support. W.G. was supported by the NSFC under Grant No.~11775021 and No.~11734002. A.W.S. was supported by the NSF under Grant
No.~DMR-1710170 and by the Simons Foundation. P.S. acknowledges financial support from the Ministry of Education via Grant No. MOE2015-T1-001-056. W.G. and P.S. would also
like to thank Boston University's Condensed Matter Theory Visitors
Program for support. The simulations in this work were carried out at computing facilities of the Nanyang Technological University and the National Supercomputing Center, Singapore.
\end{acknowledgments}

\bibliography{main}

\end{document}